\documentclass[twocolumn,showpacs,preprintnumbers,amsmath,amssymb]{revtex4}
\usepackage{graphicx}
\usepackage{dcolumn}
\usepackage{bm}

\begin{document}
\title{{Near-threshold photoelectron holography beyond the strong-field approximation}}
\author{XuanYang Lai$^{1}$, ShaoGang Yu$^{1,2}$, YiYi Huang$^{1,2}$, LinQiang Hua$^{1}$, Cheng Gong$^{1}$, Wei Quan$^{1}$}
\author{C. Figueira de Morisson Faria$^{3}$}\email{c.faria@ucl.ac.uk}
\author{XiaoJun Liu$^{1}$}\email{xjliu@wipm.ac.cn}
\affiliation{$^{1}$State Key Laboratory of Magnetic Resonance and
Atomic and Molecular Physics and Center for Cold Atom Physics, Wuhan
Institute of Physics and Mathematics, Chinese Academy of Sciences,
Wuhan 430071, China\\$^{2}$ School of Physics, University of Chinese
Academy of Sciences, Beijing 100080, China \\ $^{3}$Department of
Physics and Astronomy, University College London, Gower Street,
London WC1E 6BT, United Kingdom}

\date{\today}
\begin{abstract}

We study  photoelectron angular distributions (PADs) near the
ionization threshold with a newly developed Coulomb quantum-orbit
strong-field approximation (CQSFA) theory. The CQSFA simulations
present an excellent agreement with the result from time-dependent
Schr\"odinger equation method. We show that the low-energy
fan-shaped structure in the PADs corresponds to a subcycle
time-resolved holographic structure and stems from the significant
influence of the Coulomb potential on the phase of the
forward-scattering electron trajectories, which affects different
momenta and scattering angles unequally. For the first time, our
work provides a direct explanation of how the fan-shaped structure
is formed, based on the quantum interference of direct and
forward-scattered orbits.

\end{abstract}

\pacs{32.80.Rm, 32.80.Qk, 42.50.Hz}

\maketitle

Quantum interference of matter waves lies at the heart of quantum
mechanics. When an atom or a molecule interacts with a strong laser
field, the bound electron may be ionized by tunneling through the
barrier formed by the Coulomb potential and the laser electric field
\cite{Keldysh}. The electron wavepackets ionized at different times
with the same final momentum will interfere with each other
\cite{Becker2002AdvAtMolOptPhys}. This results in rich interference
patterns in the above-threshold ionization (ATI) photoelectron angular distributions (PADs)
\cite{paulus2005PRL}, which have been taken as an important tool in
exploring the structure and the dynamics of
atoms and molecules with
attosecond temporal resolution and angstrom spatial resolution
\cite{Meckel2014NatPhy,Xie2015PRL}.

Recently, a new type of wavepacket interference, i.e., photoelectron
holography
\cite{Huismanset2012PRL,Huismanset2010Science,Li2015SciRep,Meckel2008Science},
has provided a novel avenue for ultrafast studies of  structural and
dynamical information about the atomic or molecular medium. By
analogy with optical holography \cite{Gabor1948Nature},  the
electron wavepacket which directly drifts to the detector after
tunneling ionization is taken as reference wave, while the electron
wavepacket which further interacts with the core and then drifts to
the detector acts as signal wave. These two paths with the same
final momentum interfere with each other, forming the holographic
patterns in the   PADs. Since the signal wave scatters off the
target and encodes its structure, the hologram stores spatial and
temporal information about the core- and electron dynamics. For
example, a ``fork"-like holographic structure was experimentally
observed in the PADs of metastable xenon atoms
\cite{Huismanset2012PRL,Huismanset2010Science}. This specific
structure is produced by the direct and the laser-driven
forward-scattering electron wavepackets from the same quarter cycle
of the laser pulse; thus, subcycle time resolution is encoded in the
holographic patterns \cite{Huismanset2012PRL,Huismanset2010Science}.
Furthermore, signal and reference  waves can be born in different
quarter cycles, leading to different holographic structures
\cite{Bian2011PRA,Bian2012PRL,Bian2014PRA,Hickstein2012PRL}. For instance, a
fishbone-like holographic structure from the interference by the
direct and the backscattered electron wavepackets has been
identified experimentally \cite{Haertelt2016PRL}. This structure has
been proposed as a particularly sensitive probe of the molecular
structures \cite{Bian2012PRL}. Hence, how to decode the structural
and dynamical information about the target from a given holographic
structure has also attracted great attention. This has led to a
novel approach for  extracting the phase of the scattering amplitude
of the signal wave, providing time-resolved imaging of ultrafast
processes \cite{Zhou2016PRL}.

Nonetheless, the understanding of time-resolved photoelectron
holography is still quite preliminary. Holographic patterns are
usually understood  within the strong-field approximation (SFA)
\cite{Keldysh,Becker2002AdvAtMolOptPhys}, or  semicalssical models
in which the influence of the ionic Coulomb potential on the
dynamics of the ionized electron is fully neglected
\cite{Bian2011PRA,Bian2012PRL}. Recently, however, the Coulomb
potential has been found to play an important role in the
photoelectron spectra, leading to, e.g., an unexpected low-energy
structure
\cite{PRL2009Quan,NatPhys2009Blaga,Becker2014JPB,Wu2012PRL,Becker2015JPB}
and even a zero-energy structure \cite{Dura2013SciRep,Wei2016SciR}.
The Coulomb potential also modifies the holographic patterns,
resulting in, e.g., the reduced fringe spacing in the ``fork"-like
holographic structure \cite{Huismanset2012PRL,Huismanset2010Science}
and the appearance of the clear backscattering holography due to the
Coulomb focusing  \cite{Bian2011PRA,Bian2012PRL,Bian2014PRA}. The
physics behind the Coulomb effects is however poorly understood,
which greatly hinders a comprehensive understanding of photoelectron
holography and its potential applications in strong field and
attosecond physics.

Another structure caused by the interplay between the Coulomb
potential and the laser field is a fan-shaped interference pattern
that appears in two-dimensional PADs near the ionization threshold.
This structure has been measured in several experiments
\cite{Rudenko2004JPB,Maharjan2006JPB,Huismanset2010Science} and has
been the topic of theoretical studies since the past decade
\cite{Arbo2008,Arbo2006PRL,Chen2006PRA,Arbo2010PRA}.
Regardless, there is no direct explanation of how this pattern
forms. Empirical rules for predicting the number of fringes have
been given in \cite{Arbo2006PRL,Chen2006PRA}, but this rule loses
its efficacy as the laser intensity is increased
\cite{Marchenko2010JPB}. Furthermore, in \cite{Arbo2006PRL} the
patterns were related to laser-dressed Kepler hyperbolae with
neighboring angular momenta. However, the arguments in
\cite{Arbo2006PRL} are backed by classical-trajectory Monte-Carlo
computations, for which quantum interference is absent. This means
that there is no direct evidence that the fanlike structure can be
reproduced, or of how it develops. Subsequently, the fan-shaped
structure is reproduced with Coulomb-Volkov approximation
\cite{Arbo2008}, for which the influence of the Coulomb potential is
included in the final electron state, but not in the continuum
propagation.  Hence, it does not provide information on how the
Coulomb potential changes the electron trajectories and only allows
a vague explanation for how the patterns form. Therefore, a new
theoretical method is required to reveal the underlying physics of
the Coulomb effect on the fan-shaped interference patterns.

In this paper, we study the above-mentioned fan-shaped structure
with a Coulomb quantum-orbit strong-field approximation (CQSFA)
theory \cite{Lai2015PRA}. This newly developed approach exhibits a
very good agreement with the result from   the time-dependent
Schr\"odinger equation (TDSE), and allows a direct assessment of
quantum interference in terms of a few electron trajectories and
their phase differences.  We perform a detailed analysis of how the
fan-shaped pattern forms, and, more importantly,  show that it
corresponds to a subcycle time-resolved holographic structure
arising from the interference between the direct and the
forward-scattered electron wavepackets. This type of forward
scattering is absent in the SFA, and corresponds to trajectories
along which the electron is deflected by the Coulomb potential
without undergoing a hard collision with the core. Due to the
Coulomb potential, the phase associated with the forward-scattering
trajectories is significantly changed. These distortions are angle
dependent, and more dramatic for lower-energy photoelectrons,
resulting in the specific fan-shaped structure. Thus, our work for
the first time explains the underlying physics of the Coulomb effect
on the fan-shaped structure. Additionally, we analyze the electron
ionization dynamics and identify a clear signature of nonadiabatic
tunneling.

The CQSFA theory~\cite{Lai2015PRA} employed in this work describes
ionization in terms of quantum orbits from the saddle-point
evaluation of the ionization amplitude. Conceptually, it differs
from the Coulomb-corrected SFA (CCSFA) theory
\cite{Yan2010PRL,Yan2012PRA} and the Eikonal Volkov approximation
(EVA) \cite{Smironova2008PRA}, which are the most widespread
Coulomb-corrected strong-field approaches. While the EVA is derived
from a laser-dressed Wentzel-Kramers-Brillouin (WKB) approach in the
limit of small scattering angles, and the CCSFA constructs its
trajectories recursively starting from the Coulomb-free trajectories
used in the SFA, the CQSFA is derived using path-integral methods,
which are applied to the full time-evolution operator. From the
implementation viewpoint, there are also differences as the CCSFA
solves the direct problem of seeking the final momentum for a given
initial momentum, while the CQSFA focuses on the inverse problem. An
important consequence is that sampling in the CCSFA is implemented
to obtain a large number of orbits and then these
trajectories are binned according to their final momenta. Thus, in practice, there is a huge amount of electron
trajectories in the CCSFA, while, in the CQSFA, a few electron
trajectories suffice. For example, for each photoelectron in the
low-energy region, three trajectories within a driving-field cycle
are needed for obtaining converged PADs. Therefore, by analyzing the
phase difference between these few orbits, we can directly
understand how the interference patterns are formed and how the
Coulomb potential influences this interference.

Briefly, in the CQSFA theory, the initial state is a bound state $
\left\vert \psi _{0} (t_0)\right\rangle= e^{iI_pt_0}\left\vert \psi
_{0}\right\rangle$, and the final state is a continuum state $
|\psi_{\textbf{p}_f}(t)\rangle$ with momentum $\mathbf{p}_f$. This
gives the ionization amplitude  (in atomic units)
\cite{Becker2002AdvAtMolOptPhys}
\begin{equation}\label{Mpdir}
M(\mathbf{p}_f)=-i \lim_{t\rightarrow \infty} \int_{-\infty }^{t }d
t_0 \left\langle \psi_{\textbf{p}_f}(t) |\hat{U}(t,t_0)\hat{H}_I(t_0)| \psi
_0(t_0)\right\rangle,
\end{equation}
where $\hat{U}(t,t_0)$ is the time-evolution operator of the Hamiltonian $
\hat{H}(t)=\hat{\mathbf{p}}^{2}/2+V(\hat{\mathbf{r}})+\hat{H}_I(t)$
with  $\hat{H}_I(t)=-\hat{\mathbf{r}}\cdot \mathbf{E}(t)$. Note that
Eq.~(\ref{Mpdir}) is formally exact. Employing the Feynman
path-integral formalism \cite{Kleinert2009,Milosevic2013JMP} and the
saddle-point approximation \cite{Kopold2000OC,Carla2002PRA},
Eq.~(\ref{Mpdir}) 
becomes
\begin{eqnarray}\label{MpPathSaddle}
M(\mathbf{p}_f) & \propto &  -i \lim_{t\rightarrow \infty }
\sum_{s}\bigg\{\det \bigg[  \frac{\partial\mathbf{p}_s(t)}{\partial
\mathbf{r}_s(t_{0,s})} \bigg] \bigg\}^{-1/2} e^{i
S(\mathbf{\tilde{p}}_s,\textbf{r}_s,t_{0,s},t))} \nonumber \\ & &
\times \mathcal{C}(t_{0,s})  \left\langle
\mathbf{p}_s(t_{0,s})+\mathbf{A}(t_{0,s})\right. |\hat{H}_I
(t_{0,s})|\left. \psi _{0}\right\rangle, \,
\end{eqnarray}
where the term $ \label{eq:sfaamp} \mathcal{C}(t_0)$ is the
prefactor,  $\partial\mathbf{p}(t)/\partial \mathbf{r}(t_0)$ is
related to the stability of the trajectory,
$S(\mathbf{\tilde{p}},\textbf{r},t_0,t) =I_p t_0-
 \int_{t_0}^{t} d\tau [\mathbf{\dot{p}}\cdot
\textbf{r}(\tau)+\mathbf{\tilde{p}}^{2}/2+V(\mathbf{r})]$ denotes
the action, in which the term $\mathbf{\dot{p}}\cdot
\textbf{r}(\tau)$ is important for obtaining correct interference patterns \cite{Shvetsov-Shilovski2016PRA}, $I_p$ is the ionization potential,  $\mathbf{p}$ is
the field-dressed momentum and
$\mathbf{\tilde{p}}=\mathbf{p}+\mathbf{A}(\tau)$, with $t_0<\tau<t$,
is the electron velocity. The index $s$ denotes the different
orbits from three saddle-point equations: $
\label{tun_time_ccsfa} [ \textbf{p}_0+\textbf{A}(t_0)]^{2}/2+I_{p}=0
$, $ \textbf{\.{p}}(\tau)= -\nabla_\textbf{r}V[\textbf{r}(\tau)]$
and $ \textbf{\.r}(\tau)= \textbf{p}(\tau)+\textbf{A}(\tau)$, which
are solved using an iteration scheme  for any given final momentum
\cite{Lai2015PRA} with the assumption that  the electron is ionized
by tunneling from  $t_0$ to  $t_0^R=\text{Re}[t_0]$ and then moves
to the detector with the real time
\cite{Popruzhenko2008JMO,Popruzhenko2014JPB} (see the
supplemental material).

\begin{figure*} [tb]
\includegraphics[width=6 in]{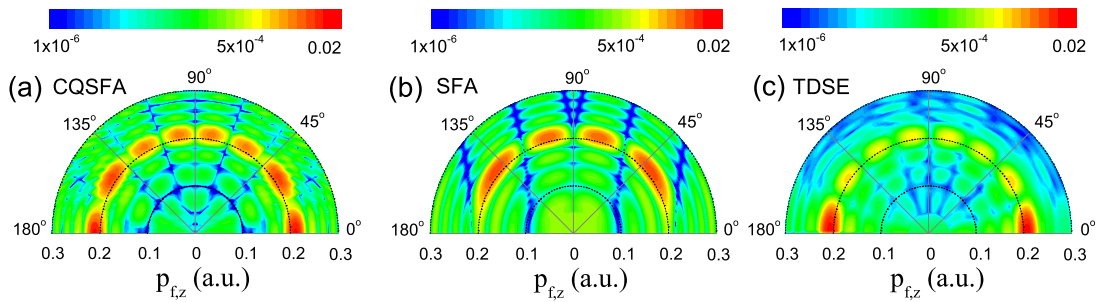}\\
\caption{(Color online) Two-dimensional PADs of  hydrogen atom
($I_p=0.5$ a.u.) near the ionization threshold in a linearly
polarized laser field of intensity $I=2\times 10^{14}$ W/cm$^2$ and
wavelength $\lambda=800$ nm, for momenta $p_f<0.3$ a.u. Panels (a),
(b) and (c) refer to CQSFA, SFA, and TDSE, respectively. The
momentum component along the laser polarization direction is given
by $p_{f,z}$.  In panels (a) and (b), we have used  $\textbf{E}(t)=
\hat{\textbf{z}}E_0 \sin \omega t$  over five cycles, while in panel
(c) we have taken a long laser pulse $\textbf{E}(t)=\hat{\textbf{z}}
E_{0} \sin\omega t \times f(t) $ with a trapezoidal profile $f(t)$
(up and down-ramped over 2 cycles, constant over 8 cycles). The TDSE
spectra have been computed employing a window operator of width
$5\times10^{-4}$ a.u. as discussed in \cite{qprop}.  All panels have
been normalized to the same range to facilitate a direct comparison.
} \label{fig1}
\end{figure*}

Figs.~\ref{fig1}(a) and (b) exhibit the two-dimensional PADs near
the ionization threshold computed for the hydrogen atom in a
linearly polarized laser field  with the CQSFA and the SFA,
respectively.  As a benchmark, we take the \emph{ab initio} TDSE
calculation shown in Fig.~\ref{fig1}(c), which is solved using the
freely available software Qprop \cite{qprop}. Our results show
significant changes in the PAD for the CQSFA, in comparison with the
SFA simulation. Indeed, there are eight clear peaks in the first ATI
ring at the momentum $p_f \sim 0.2 $ a.u. \cite{Arbo2010PRA}, while
only four peaks are found in the corresponding SFA simulations. Note
that the number of the peaks will be changed for other initial
states of atom and  laser parameters; for more details, see the
supplemental material. Moreover, in Fig.~\ref{fig1}(a) a clear
radial fanlike pattern is present between the threshold region and
the onset of the ATI ring, which completely disappears in
Fig.~\ref{fig1}(b). The overall interference pattern in CQSFA
exhibits a very good agreement with the TDSE simulations in
Fig.~\ref{fig1}(c), reflecting the significant role of the Coulomb
potential on  strong-field ionization, which is consistent with
previous publications \cite{Arbo2008,Arbo2006PRL}. There is however
a slight quantitative discrepancy between CQSFA and TDSE in the
amplitudes of the eight peaks. Possible reasons are provided in the
supplemental material. In this work, however, we focus on how the
Coulomb potential leads to the generation of the fanlike structure
by analyzing quantum orbits from the CQSFA.

\begin{figure} [tb]
\includegraphics[width=3.5 in]{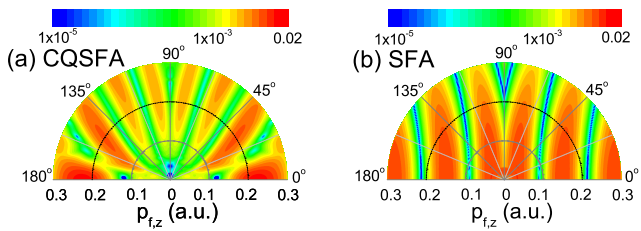}\\
\caption{(Color online) Same as Figs.~\ref{fig1}(a) and (b), but
calculated with quantum orbits occurred in one optical cycle.  }
\label{fig2}
\end{figure}

In Fig.~\ref{fig2} we consider only the interference of the orbits
within one driving-field cycle, i.e., the intracycle interference
\cite{Arbo2010PRA}. The CQSFA outcome [Fig.~\ref{fig2}(a)] exhibits
eight interference stripes, which, near the ionization threshold,
roughly point to zero momentum, showing a divergent structure. In
contrast, for the SFA simulations in Fig.~\ref{fig2}(b), there are
only six approximately vertical interference stripes, which bend as
the transverse momentum increases. If quantum-orbit contributions
from other optical cycles are also added coherently, intercycle
interference \cite{Arbo2010PRA} forms characteristic ATI rings
centered around zero momentum. The modulation between the intracycle
and intercycle interferences results in the clear eight peaks in the
first ATI ring in  CQSFA and the four ATI peaks in   SFA, just as
illustrated in Figs.~\ref{fig1}(a) and (b), respectively. Moreover,
the divergent structure in Fig.~\ref{fig2}(a) corresponds to the
fanlike pattern shown in the PADs. Thus, the difference between
Figs.~\ref{fig1}(a) and (b) stems from the influence of the Coulomb
potential on the intracycle interference.

\begin{figure} [tb]
\includegraphics[width=3.5 in]{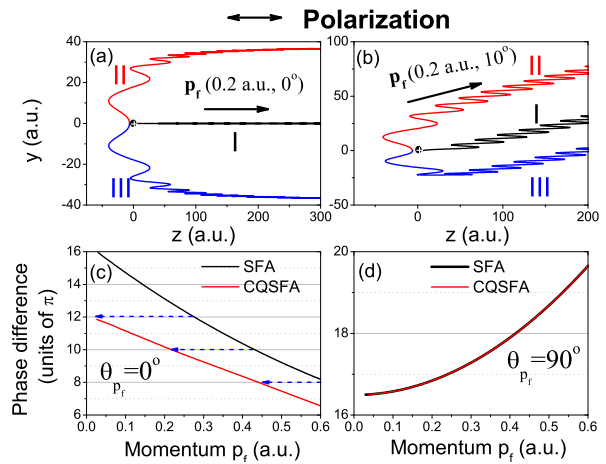}\\
\caption{(Color online) (a) and (b) Illustration of three orbits
from the CQSFA theory in the $yz$ plane for electrons with fixed
final momentum $p_f=0.2$ a.u. along the $0^{\circ}$ and $10^{\circ}$
directions with respect to the laser polarization, respectively.
The laser polarization is along the $z$ axis. The black circle at
the position $(0,0)$ denotes   the nucleus. (c) and (d) Phase
difference  between  orbits I and II as a function of $p_f$ along
$0^{\circ}$ and $90^{\circ}$ directions with respect to the laser
polarization, respectively. The blue arrows denote the shift of the
positions of the interference maxima. } \label{fig3}
\end{figure}

More insight can be gained by analyzing the positions of the
interference stripes in the PADs. In the CQSFA theory, the amplitude
of the intracycle interference in the low-energy region is mainly
determined by three quantum orbits \cite{Lai2015PRA}.
Figs.~\ref{fig3}(a) and (b)  depict these orbits in the $yz$ plane
for electrons with fixed final momentum $p_f=0.2$ a.u. along the
$0^{\circ}$ and $10^{\circ}$ directions with respect to the laser
polarization, respectively. For orbit I,  the electron moves
directly towards the detector without returning to the parent ion.
In contrast, for   orbits II and III, the electron will turn around
the core and then move to the detector along Kepler hyperbolae to
which a quiver motion caused by the laser field is superimposed
\cite{Arbo2006PRL,Yan2010PRL}. Therefore, the patterns in
Fig.~\ref{fig2}(a) correspond to a holographic structure from the
interference between the direct trajectories and
forward-scattering trajectories, which are deflected by the
core but do not undergo hard collisions. Orbits I and II are similar
to the so-called short and long trajectories in   SFA
\cite{Bian2011PRA}, while orbit III is not found in the SFA and is
observed after the Coulomb potential is considered
\cite{Yan2010PRL,Huismanset2010Science}. If the final momentum is
along the laser polarization [Fig.~\ref{fig3}(a)], orbits II and III
are symmetric with respect to the polarization direction.  With
increasing scattering angle $\theta_{\mathbf{p}_f}$,  orbit III will
experience a stronger attraction from the Coulomb potential, leading
to a larger deflection [see, e.g., Fig.~\ref{fig3}(b)]. Due to
Coulomb defocusing, the  amplitude   of orbit III decreases
significantly with increasing scattering angle, while the
contributions from orbits I and II become dominant. This results in
well defined reference and probe signals in the holographic
patterns: orbits I and II, respectively.

Hence, we will focus on the phase difference between orbits I and
II, $\Delta\Phi=\Phi_{\text{I}}-\Phi_{\text{II}}$, which is directly
related to the interference pattern and  is displayed in
Figs.~\ref{fig3}(c) and (d) as a function of the final electron
momentum, for parallel and perpendicular scattering angles,
respectively.  A similar analysis has been employed in our previous
publications \cite{Lai2013PRA,Lai2015PRA}. For
$\theta_{\mathbf{p}_f}=0^{\circ}$, the phase difference decreases if
the Coulomb potential is incorporated. This shifts the interference
maxima in the CQSFA spectra towards lower energies, in comparison
with their SFA counterparts (see the blue dashed arrows).
Physically, this happens because, in comparison with orbit I, orbit
II accumulates a larger positive phase contribution from the Coulomb
potential as it passes by the core, $-\int_{t'}^t
V[\textbf{r}(\tau)] d\tau$ [see Eq.~(\ref{MpPathSaddle})]. With
increasing $\theta_{\mathbf{p}_f}$, the above-mentioned shift
becomes smaller and is almost negligible for perpendicular emission
[Fig.~\ref{fig3}(d)].  This is not surprising since  for
$\theta_{\mathbf{p}_f}=90^{\circ}$ orbits I and II  are symmetric
with respect to the $y$ axis. Thus, the influence of the Coulomb
potential on the two orbits is the same. A similar result has been
reported in the study of the interference carpets in ATI
\cite{Korneev2012PRL}. Therefore, more interference stripes will
appear in the low-energy region for the CQSFA, in agreement with
Fig.~\ref{fig2}(a). Furthermore, the shift in $\Delta\Phi$ is more
significant for smaller momenta [Fig.~\ref{fig3}(c)] as, in this
case, the electron will need a longer time to leave the core region.
This will result in a larger positive  phase contribution from the
Coulomb potential and  a larger decrease in $\Delta\Phi$. Therefore,
the interference maxima in the photoelectron spectra will shift more
dramatically for smaller momenta. A similar behavior is observed for
other emission angles $\theta_{\mathbf{p}_f}$, leading to the
fanlike structures in Figs.~\ref{fig1}(a) and \ref{fig2}(a).

\begin{figure} [tb]
\includegraphics[width=3.5 in]{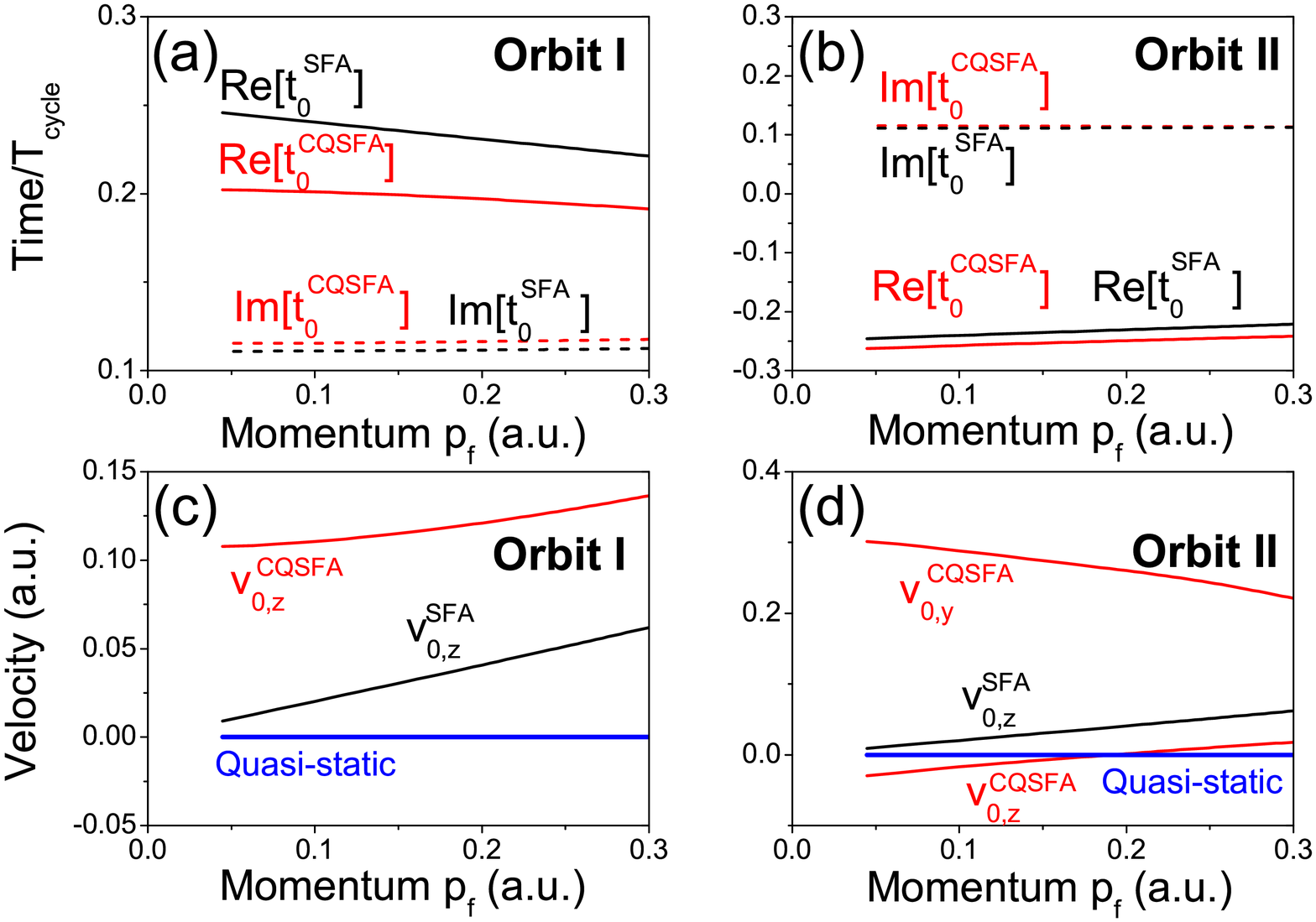}\\
\caption{(Color online) (a) and (b) The real part (solid lines) and
imaginary part (dashed lines) of the tunneling time  for electrons
with $\theta_{\mathbf{p}_f}=0^{\circ}$ in CQSFA (red)  and  in SFA
(black). (c) and (d) The corresponding initial velocity  at the
tunneling exit. The quasi-static expected velocity is denoted by the
blue lines. (a)(c): orbit I and (b)(d): orbit II.   } \label{fig4}
\end{figure}

Finally, Figs.~\ref{fig4}(a) and (b) show the tunneling time of the
photoelectrons  from   CQSFA and  SFA  for
orbits I and II, respectively, for a specific angle
$\theta_{\mathbf{p}_f}=0^{\circ}$. Due to the ionization by
tunneling, the time $t_0$ becomes complex
\cite{Becker2002AdvAtMolOptPhys}, and $\mathrm{Im}[t_0]>0$ can be
related to the tunneling time through the potential barrier. For
each kind of orbit, the photoelectrons are initially ionized within
a temporal window of about $0.02T_{cycle}$ ($\sim60$ attoseconds),
and the orbits I and II originate from the adjacent quarter cycles
of the laser pulse. Therefore, the subcycle fan-shaped structure
has recorded attosecond time-resolved electronic dynamics. In
comparison with the SFA, for a given final momentum $\textbf{p}_f$,
$\text{Re}(t_0)$ in the CQSFA approaches the driving-field crossing
($ t=0$) or its crest ($t=-0.25T_{\text{cycle}}$) for orbit I or II,
respectively. This increases (decreases) the initial field-dressed
momentum $\mathbf{p}_0$ for orbit I (orbit II). Along orbit I, the
electron compensates the deceleration in the Coulomb potential as it
moves towards the detector, while, for orbit II, the electron is
accelerated significantly along the polarization direction due to
the interplay of the Coulomb potential and the laser field
\cite{Popruzhenko2008JMO,Lai2015PRA}. In Figs.~\ref{fig4}(c) and (d)
we illustrate the change in   the initial velocity
$\textbf{v}_0=\textbf{p}_0+\textbf{A}(\text{Re}[t_0])$ at the
tunneling exit. Both CQSFA and SFA simulations significantly deviate
from the adiabatic tunneling theory (blue lines in Fig.~\ref{fig4}),
in which the electron is assumed to begin its journey in the
continuum with vanishing velocity
\cite{Keldysh,Ivanov2005JMO,NatPhys2015Pedatzur}.

In summary, we have performed a detailed analysis of the low-energy
fanlike structure observed in PADs using a CQSFA theory, in which
only a few electron trajectories are required to describe
strong-field ionization, and which poses no restriction upon the
scattering angle. We show that this structure constitutes a subcycle
time-resolved holographic pattern from the interference of direct
electron trajectories and forward-scattered trajectories that are
deflected, but do not undergo hard collisions with the core. We go
beyond existing studies by providing direct and in-depth evidence of
how the Coulomb potential alters the phase of the forward-scattering
trajectories, which affects different scattering angles and electron
momenta unequally, leading to the above-mentioned fanlike structure.
The present method can be applied to the understanding of Coulomb
effects on other holographic patterns, e.g.,  the well-known reduced
fringe spacing in the ``fork"-like holographic structure
\cite{Huismanset2012PRL,Huismanset2010Science}.

We thank Prof.~Wilhelm Becker and Prof.~Xue-Bin Bian for many useful
discussions. This work is supported by the National Basic Research
Program of China Grant (No. 2013CB922201), the NNSF of China (Nos.
11374329, 11334009, 11474321, and 11527807) and the UK EPSRC (No.
EP/J019240/1).


\begin{thebibliography}{99}

\bibitem{Keldysh} L. V. Keldysh, Zh. Eksp. Teor. Fiz. \textbf{47}, 1945 (1964) [Sov.
Phys. JETP \textbf{20}, 1307 (1965)]; F. H. M. Faisal, J. Phys. B \textbf{6}, L89
(1973); H. R. Reiss, Phys. Rev. A \textbf{22}, 1786 (1980).

\bibitem{Becker2002AdvAtMolOptPhys} W. Becker, F. Grasbon, R. Kopold, D. B. Milo\v{s}evi\'c, G. G. Paulus and H. Walther, Adv. At. Mol. Opt. Phys. \textbf{48}, 35 (2002).

\bibitem{paulus2005PRL} F. Lindner \emph{et. al}., Phys. Rev. Lett. \textbf{95}, 040401 (2005).

\bibitem{Meckel2014NatPhy} M. Meckel, A. Staudte, S. Patchkovskii, D. M. Villeneuve,
P. B. Corkum, R. D\"orner, and M. Spanner, Nat. Phys. \textbf{10},
594 (2014).
\bibitem{Xie2015PRL} X. H. Xie,  Phys. Rev. Lett. \textbf{114}, 173003 (2015).

\bibitem{Huismanset2010Science} Y. Huismans \emph{et al.}, Science \textbf{331}, 61 (2010).


\bibitem{Huismanset2012PRL} Y. Huismans \emph{et al.}, Phys. Rev. Lett. \textbf{109}, 013002 (2012).

\bibitem{Meckel2008Science} M. Meckel \emph{et al.}, Science \textbf{320}, 1478 (2008).




\bibitem{Li2015SciRep}M. Li, X. Sun, X. Xie, Y. Shao, Y. Deng, C. Wu, Q. Gong,
and Y. Liu, Sci. Rep. \textbf{5}, 8519 (2015).

\bibitem{Gabor1948Nature} D. Gabor, Nature (London) \textbf{161}, 777 (1948).


\bibitem{Bian2011PRA}X.-B. Bian\emph{ et al.}, Phys. Rev. A \textbf{84}, 043420 (2011).
\bibitem{Bian2012PRL}X.-B. Bian and A. D. Bandrauk, Phys. Rev. Lett. \textbf{108}, 263003 (2012).
\bibitem{Bian2014PRA}X.-B. Bian and A. D. Bandrauk, Phys. Rev. A \textbf{89}, 033423 (2014).
\bibitem{Hickstein2012PRL} Daniel D. Hickstein  \emph{et al.}, Phys. Rev. Lett. \textbf{109}, 073004 (2012).




\bibitem{Haertelt2016PRL} M. Haertelt, X.-B. Bian, M. Spanner, A. Staudte, and P. B. Corkum, Phys. Rev. Lett. \textbf{116}, 133001 (2016).


\bibitem{Zhou2016PRL}Y. M. Zhou, O. I. Tolstikhin, and T. Morishita, Phys. Rev. Lett. \textbf{116}, 173001 (2016).

\bibitem{NatPhys2009Blaga} C. I. Blaga, F. Catoire, P. Colosimo, G. G. Paulus, H. G. Muller, P. Agostini, and L. F. Dimauro, Nat. Phys. \textbf{5}, 335 (2009).

\bibitem{PRL2009Quan} W. Quan \emph{et al.}, Phys. Rev. Lett.  \textbf{103}, 093001 (2009).


\bibitem{Becker2014JPB} W. Becker et al., J. Phys. B \textbf{47}, 204022 (2014).

\bibitem{Wu2012PRL}    C. Y. Wu \emph{et al.}, Phys. Rev. Lett. \textbf{109}, 043001 (2012).
\bibitem{Becker2015JPB} W. Becker and  D. B. Milo\v{s}evi\'c, J. Phys. B \textbf{48}, 151001 (2015).
\bibitem{Dura2013SciRep} J. Dura \emph{et al.}, Sci. Rep. \textbf{3}, 2675 (2013).
\bibitem{Wei2016SciR} W. Quan \emph{et al.},  Sci. Rep. \textbf{6}, 27108 (2016).



\bibitem{Rudenko2004JPB} A. Rudenko \emph{et al.}, J. Phys. B \textbf{37}, L407 (2004).

\bibitem{Maharjan2006JPB} C. M. Maharjan \emph{et al.}, J. Phys. B \textbf{39}, 1955 (2004).

\bibitem{Arbo2006PRL} D. G. Arb\'o \emph{et al}.,  Phys. Rev. Lett. \textbf{96}, 143003 (2006);  D. G. Arb\'o \emph{et al}.,  Phys. Rev. A \textbf{78}, 013406 (2008).




\bibitem{Arbo2008} D. G. Arb\'o \emph{et al}., Phys. Rev. A \textbf{77}, 013401 (2008).



\bibitem{Chen2006PRA} Z. Chen \emph{et al.}, Phys. Rev. A 74, 053405 (2006).

\bibitem{Arbo2010PRA}   D. G. Arb\'o \emph{et al}., Phys. Rev. A \textbf{81}, 021403(R) (2010);  D. G. Arb\'o \emph{et al}., Nucl. Instrum. Methods B \textbf{279}, 24 (2012).



\bibitem{Marchenko2010JPB} T. Marchenko, H. G. Muller, K. J. Schafer and M. J. J. Vrakking,  J. Phys. B  \textbf{43}, 095601 (2010).

\bibitem{Lai2015PRA} X. Y. Lai, C. Poli, H. Schomerus, and C. Figueira de Morisson Faria, Phys. Rev. A \textbf{92}, 043407 (2015).


\bibitem{Yan2010PRL} T. M. Yan, S. V. Popruzhenko, M. J. J. Vrakking, and D. Bauer,
Phys. Rev. Lett. \textbf{105}, 253002 (2010).
\bibitem{Yan2012PRA} T. M. Yan and D. Bauer, Phys. Rev. A \textbf{86}, 053403 (2012).

\bibitem{Smironova2008PRA}   O. Smirnova, M. Spanner, and M. Ivanov, Phys. Rev. A \textbf{77}, 033407 (2008).

\bibitem{Kleinert2009} H. Kleinert,  \emph{Path integrals in quantum mechanics, statistics, polymer physics,
and financial markets}, (World Scientific, 2009).

\bibitem{Milosevic2013JMP} D. B. Milo\v{s}evi\'c, J. Math. Phys. \textbf{54}, 042101(2013).



\bibitem{Carla2002PRA} C. Figueira de Morisson Faria, H. Schomerus, and W.
Becker, Phys. Rev. A \textbf{66}, 043413 (2002).

\bibitem{Kopold2000OC} R. Kopold, W. Becker, and M. Kleber, Opt.
Commun. \textbf{179}, 39 (2000).
\bibitem{Shvetsov-Shilovski2016PRA}  N. I. Shvetsov-Shilovski \emph{et al}., Phys. Rev. A \textbf{94}, 013415
(2016).

\bibitem{Popruzhenko2008JMO} S. V. Popruzhenko and D. Bauer, J. Mod. Opt. \textbf{55}, 2573
(2008).

\bibitem{Popruzhenko2014JPB} S. V. Popruzhenko, J. Phys. B \textbf{47}, 204001 (2014).

\bibitem{qprop}  D. Bauer and P. Koval, Comput. Phys. Commun. \textbf{174}, 396
(2006); see also www.qprop.de.



\bibitem{Lai2013PRA}   X. Y. Lai and  C. Figueira de Morisson Faria, Phys. Rev. A \textbf{88}, 013406 (2013).
\bibitem{Korneev2012PRL}   Ph. A. Korneev \emph{et al.}, Phys. Rev. Lett. \textbf{108},
223601 (2012).



\bibitem{NatPhys2015Pedatzur} O. Pedatzur \emph{et al}., Nat. Phys. \textbf{11}, 815 (2015).


\bibitem{Ivanov2005JMO}   M. Y. Ivanov, M. Spanner, and O. Smirnova, J. Mod. Opt. \textbf{52}, 165 (2005).
%
%
%
%
%
%
%
%
%
%
%
%
%
%
%
%
%
%
%
%
%
\end{thebibliography}
\end{document}